\documentclass[11pt]{article}
\usepackage{amsmath}
\usepackage{amsfonts}

\usepackage{amssymb,graphicx}

\usepackage{amssymb}

\def\bo{o^\dagger}
\def\tD{\tilde{\Delta}}\def\tth{\tilde{\Theta}}
\def\be{\begin{equation}}
\def\ee{\end{equation}}

\begin{document}

\title{One-sided type-D Ricci-flat metrics}
\author{Paul Tod\footnote{email: tod@maths.ox.ac.uk }\\Mathematical Institute,\\Oxford University}
%

\maketitle

\begin{abstract}
We consider four-dimensional, Riemannian, Ricci-flat metrics for which one or other of the self-dual or anti-self-dual Weyl tensors is type-D. Such metrics always have a valence-2 Killing spinor, and therefore a Hermitian structure and at least one Killing vector. We rederive the results of Przanowski and collaborators, that these metrics can all be given in terms of a solution of the $SU(\infty)$-Toda field equation, and show that, when there is a second Killing vector commuting with the first, the method of Ward can be applied to show that the metrics can also be given in terms of an axisymmetric solution of the flat three-dimensional Laplacian. Thus in particular the field equations linearise.

As a corollary, we show that the same technique linearises the field equations for a four-dimensional Einstein metric with anti-self-dual Weyl tensor and two commuting symmetries.

We reduce the Einstein equations with non-zero scalar curvature and one-sided type-D Weyl tensor, excluding the K\"ahler-Einstein case, to a modified but not integrable Toda equation.

Some examples of the constructions are given.

\end{abstract}
\section{Introduction}

In this note we consider four-dimensional, Riemannian, Ricci-flat metrics for which one or other of the self-dual or anti-self-dual Weyl tensors is type-D in the Petrov-Pirani-Penrose classification (for which see e.g. \cite{pr}). We'll call these metrics \emph{one-sided type-D}, a term in use at least since 1984 (see \cite{pb}). Our motivation comes indirectly from the Chen-Teo metric \cite{ct}. This has two commuting Killing vectors and is Riemannian and Ricci-flat and was obtained by inverse-scattering methods, which are available since the Einstein equations in this case are known to be completely integrable, (see e.g. \cite{bv}). It was discovered by Aksteiner \cite{ak} that the Chen-Teo metric is one-sided type-D, and therefore, as we shall see, Hermitian. In this note we ask: can one do more with it, in the light of the general solution of one-sided type-D vacuum metrics in \cite{pb,pb2}?

In \cite{pb,pb2} the Einstein vacuum equations, subject to these restrictions, are reduced to the $SU(\infty)$-Toda equation (we'll omit the term ``$SU(\infty)$'' henceforth). It's known that the Toda equation linearises if the solution has an extra symmetry, \cite{w1}, which suggests that the field equations for one-sided type-D Ricci-flat metrics with a second symmetry commuting with the first linearise, and we shall see here that they do.

\medskip

Our method is to start in Section 2 with the assumption of a 4-dimensional Riemannian, Ricci-flat metric which is one-sided type-D, or equivalently (as we show) is Hermitian. Then we use the two-component spinor formalism to rederive the expressions of \cite{pb2} for the metric in terms of a solution $u$ of the Toda field equation (\ref{13}). The metric automatically has a Killing vector which arises from a valence-2 Killing spinor, which in turn is a consequence of the type-D-ness (by \cite{pw}), and our rederivation serves to explain the occurence of this Killing vector. Then in Section 3 we add the assumption that there is a second Killing vector and deduce that, after possible redefinitions of coordinates and $u$ preserving the Toda field equation, the second Killing vector must be a symmetry of $u$ corresponding to an ignorable coordinate which can be taken to be $y$ in (\ref{13}). Then in Section 4 we exploit the observation in \cite{w1} that solutions of the Toda field equation independent of $y$ correspond to axisymmetric solutions of the flat 3-dimensional Laplacian. We arrive at our main result: that Riemannian, Ricci-flat, one-sided type-D 4-metrics with two commuting Killing vectors are in one-to-one correspondence with axisymmetric solutions of the flat three-dimensional Laplacian. The field equations, known to be completely integrable in this case, in fact linearise. We also give a corollary: by \cite{p} (see also \cite{t1}, \cite{t2}) the general ASD Einstein metric with a symmetry and non-zero Ricci scalar can be found in terms of a solution of the Toda field equation; consequently if there is a second Killing vector commuting with the first then again by \cite{w1} the solution can be given in terms of a solution of the axisymmetric Laplace equation -- in this case too the Einstein equations linearise. In Section 5 we show that the field equations for one-sided type-D Einstein metrics reduce to a modified Toda field equation which is not integrable (\cite{F}), in Section 6 we deal with a puzzling feature of the constructions, and in Section 7 we give some examples of all the reductions.

{\bf{Acknowledgements:}} I am grateful to Dr Steffen Aksteiner of the AEI, Golm, for discussions about the Chen-Teo metric, and for telling me that it was one-sided type-D, and to the Institut Mittag-Leffler in Djursholm, Sweden for hospitality as part of the programme `General Relativity, Geometry and Analysis' during September 2019, supported by the Swedish Research Council under grant no. 2016-06596. I am grateful to Dr Maciej Dunajski of DAMTP Cambridge for references \cite{p,p2,pb,pb2} and useful discussions, in the course of which we realised that the construction given here must be possible, and I am grateful to Profs Jenya Ferapontov and Boris Kruglikov for information about integrability of the modified Toda equation (\ref{49a}).

\section{One-sided type-D}
In this section, we rederive the results of \cite{pb2} in the style of \cite{t1,t2}. The virtue of this rederivation is that one sees how it follows from the algebraic assumption on the Weyl spinor, and one also sees the origin of the first Killing vector in this assumption. For background on the 2-component spinor formalism see \cite{ht} or \cite {pr}.

\medskip

Start then with a Riemannian, Ricci-flat and one-sided type-D metric. The unprimed Weyl spinor (as a matter of convenience, call this the SD Weyl spinor - it would more properly be the ASD Weyl spinor but this makes very little difference for us) can be taken to be
\be\label{1}\psi_{ABCD}=\psi o_{(A}o_B\bo_C\bo_{D)}
\ee
with real $\psi=6\Psi_2$ in NP conventions, and the spinor normalisation
\[o_Ao^{\dagger A}=1.\]
We'll assume the metric is non-trivially type-D, in the sense that the SD Weyl spinor is not zero, so that $\psi$ is not the zero function. The argument of \cite{pw} still goes through to show that the spinor field $\omega_{AB}$ defined as
\be\label{2}\omega_{AB}=i\psi^{-1/3}o_{(A}\bo_{B)}\ee
is a (real) Killing spinor, or equivalently
\[\nabla_{A'(A}\omega_{BC)}=0.\]
Therefore
\be\label{3}
\nabla_{A'A}\omega_{BC}=\epsilon_{A(B}K_{C)A'}\ee
for some real vector $K_{AA'}$, which is then necessarily a Killing vector. To see this necessity, calculate commutators as follows (conventions following \cite{pr}): by Ricci-flat-ness
\be\label{k11}0=\Delta_{A'B'}\omega^{CD}=\nabla_{(A'}^{\;\;(C}K_{B')}^{\;\;D)},\ee
and with zero Ricci scalar
\be\label{99}\Delta_{AB}\omega_{CD}=\psi_{EABC}\omega_{D}^{\;\;E}+\psi_{EABD}\omega_{C}^{\;\;E},\ee
whence
\[\psi_{E(ABC}\omega_{D)}^{\;\;E}=0,\]
as we know (this follows from (\ref{1}), (\ref{2})), and the trace of (\ref{99}) on $BC$ gives
\[\nabla_{AA'}K_D^{\;\;A'}=\psi_{ABDE}\omega^{BE}=-\frac{1}{3}\psi\omega_{AD}.\]From the trace of this we deduce $\nabla_aK^a=0$, so with (\ref{k11}) we see that $K^a$ is a Killing vector, as claimed, and furthermore we may write
\be\label{4}
\nabla_{AA'}K_{BB'}=-\frac16\psi\omega_{AB}\epsilon_{A'B'}+\chi_{A'B'}\epsilon_{AB},\ee
for symmetric $\chi_{A'B'}$ (which we use in Section 6). The identity
\[\nabla_a\nabla_bK_c=R_{bcad}K^d,\]
which holds for any Killing vector, now gives
\[\nabla_{AA'}(-\frac16\psi\omega_{BC})=-\psi_{ABCD}K_{A'}^{\;\;D},\]
which is equivalent to
\be\label{44}\omega_{AC}K_{A'}^{\;\;C}=-\frac16\psi^{-5/3}\nabla_a\psi,\ee
and which also follows by contracting (\ref{3}) with $\omega^{BC}$, so this is an identity, and
\be\label{k4}\nabla_{AA'}\chi_{B'C'}=-\psi_{A'B'C'D'}K_A^{\;\;D'}.\ee
We find, introducing $\phi_{AB}$ as
\be\label{5}
\phi_{AB}=2io_{(A}\bo_{B)}=2\psi^{1/3}\omega_{AB},\ee
that there is an almost complex structure determined by
\be\label{6}J_a^{\;b}:=\phi_A^{\;B}\delta_{A'}^{\;B'},\ee
which is then easily seen to be integrable -- thus the metric is Hermitian by virtue of being type-D and Ricci-flat. Conversely, if there is an integrable complex structure of the form of (\ref{6}) (i.e. with this duality) then necessarily $o^A$ is geodesic and shear-free and therefore, by the Goldberg-Sachs Theorem (see e.g. \cite{pr}), is a repeated PND of the Weyl spinor. Then $o^{\dagger A}$ is another and so the Weyl spinor is type-D. Therefore the properties of being Hermitian and being one-sided type-D are here equivalent.

With (\ref{5}) we obtain
\be\label{7}\phi_{AB}K^B_{\;A'}=2\psi^{1/3}\omega_{AB}K^B_{\;A'}=\nabla_a(\psi^{-1/3}).\ee
(By a general argument, this must be a gradient as it defines the Hamiltonian for $K$).

We can proceed to find a metric ansatz as follows: set $W^{-1}:=K^bK_b$ and start an orthonormal basis of one-forms with
\[\theta^0=W^{1/2}K=W^{-1/2}(dt+\omega),\]
where $K^a\partial_a=\partial_t$ and we've introduced a presently unknown one form $\omega$. Introduce $\theta^1$ as
\[\theta^1=J\theta^0=W^{1/2}JK=W^{1/2}\phi_A^{\;B}K_{BA'}=-W^{1/2}d(\psi^{-1/3}),\]
using (\ref{7}), and so $\theta^1=W^{1/2}dz$ with $z=-\psi^{-1/3}$. Next we can choose $\theta^2,\theta^3$ orthogonal to $\theta^0,\theta^1$ and such that $J\theta^2=\theta^3$. There will be a complex coordinate $\zeta$ such that
\[\theta^2+i\theta^3=W^{1/2}e^{u/2}d\zeta,\]
for some real $u$, and the metric is
\be\label{8}g=W^{-1}(dt+\omega)^2+W(dz^2+e^ud\zeta d\overline\zeta),\ee
in terms of functions $u,W$ and the one-form $\omega$, all to be determined.

We haven't exhausted the information in integrability of the complex structure. The holomorphic one-forms are
\[e^1:=\theta^0+i\theta^1=W^{-1/2}(dt+\omega)+iW^{1/2}dz,\;\;e^2=\theta^2+i\theta^3=W^{1/2}e^{u/2}d\zeta,\]
and at once
\[e^1\wedge e^2\wedge de^2=0,\]
but the vanishing of $e^1\wedge e^2\wedge de^1$ leads to
\be\label{9}d\zeta\wedge(d\omega-idz\wedge dW)=0,\ee
which we leave for the moment (this will be part of the `monopole' equation).

From (\ref{4}) we have, written as forms
\be\label{10}dK=-\frac16\psi^{2/3}J+\mbox{  ASD terms},\ee
with $J$ the complex structure as a 2-form:
\[J=2(\theta^0\wedge\theta^1+\theta^2\wedge\theta^3).\]
However, with
\be\label{14}d\omega=\alpha dy\wedge dz+\beta dz\wedge dx+\gamma dx\wedge dy,\ee
we obtain
\[dK=d(W^{-1}(dt+\omega))=(dt+\omega)\wedge W^{-2}dW+W^{-1}d\omega\]
\[=c_1z^{-2}(\theta^0\wedge\theta^1+\theta^2\wedge\theta^3)+f_1(\theta^0\wedge\theta^1-\theta^2\wedge\theta^3)+f_2(\theta^0\wedge\theta^2-\theta^3\wedge\theta^1)+f_3(\theta^0\wedge\theta^3-\theta^1\wedge\theta^2)\]
where the $f_i$ are the coefficients of the ASD terms in (\ref{10}). From this we read off
\be\label{15}\alpha=-W_x,\;\beta=-W_y,\;\gamma=-e^u(W_z-2c_1\frac{W^2}{z^2}),\ee
and $c_1$ is a constant fixed by earlier choices (in fact $c_1=-1/3$ but it is convenient to leave it in the formulae). Note (\ref{9}) is a consequence of these, but there is a stronger integrability condition to check for (\ref{14}), namely
\be\label{15a}0=d^2\omega=(\alpha_x+\beta_y+\gamma_z)dx\wedge dy\wedge dz.\ee
We'll come back to this when we have an expression for $W$.

We next calculate the SD curvature with this choice for $d\omega$ by Cartan calculus based on a normalised triad of SD 2-forms: with an orthonormal basis of one-forms chosen as
\[\theta^0=W^{-1/2}(dt+\omega),\;\theta^1=W^{1/2}dz,\;\theta^2=W^{1/2}e^{u/2}dx,\;\theta^3=W^{1/2}e^{u/2}dy,\]
define an orthonormal basis of SD 2-forms as
\[\phi^1=\theta^0\wedge\theta^1+\theta^2\wedge\theta^3,\phi^2=\theta^0\wedge\theta^2+\theta^3\wedge\theta^1,\phi^3=\theta^0\wedge\theta^3+\theta^1\wedge\theta^2,\]
solve
\[d\phi^i=-\alpha^i_{\;j}\wedge\phi^j\]
for $\alpha^i_{\;j}$ to find
\[\alpha^1_{\;2}=C\theta^2,\;\alpha^3_{\;1}=-C\theta^3,\;\alpha^2_{\;3}=E\theta^0+G\theta^2+H\theta^3,\]
with
\[E=-c_1z^{-2}W^{1/2},\;G=\frac12W^{-1/2}e^{-u/2}u_y,\;H=-\frac12W^{-1/2}e^{-u/2}u_x,\]
and
\[C=-\frac12W^{-1/2}u_z-c_1z^{-2}W^{1/2}.\]
Now obtain the curvature components subject to Ricci flatness from
\[ \Omega^i_{\;j}=d\alpha^i_{\;j}+\alpha^i_{\;k}\wedge\alpha^k_{\;j}=  \Omega^i_{\;j\cdot k}\phi^k,\]
with $\epsilon^{ijk}\Omega_{ij\cdot k}=0$ (which encodes vanishing Ricci scalar; these indices are raised and lowered by $\delta_{ij},\delta^{ij}$), to find from $\Omega^1_{\;2}+i\Omega^1_{\;3}$ that necessarily
\[W=-\frac{z^2u_z}{2c_1}+f(z),\]
with
\[f'-2\frac{f}{z}+c_1\frac{f^2}{z^2}=0,\]
so that
\[f=\frac{z^2}{c_1z+c_2},\]
for a constant $c_2$. Then from $\Omega^2_{\;3}$ we find that $u$ satisfies the \emph{Toda field equation}:
\be\label{13}u_{xx}+u_{yy}+(e^u)_{zz}=0,\ee
and $c_1f=z$ so that $c_2=0$ and
\be\label{17} W=\frac{1}{c_1}z(1-\frac12zu_z).\ee
Now we can check the integrability condition (\ref{15a}). In fact  $G=W/z$ satisfies a familiar monopole equation:
\[G_{xx}+G_{yy}+(e^uG)_{zz}=0.\]
The SD curvature components are now
\[\Omega^1_{\;2}=c_1z^{-3}\phi^3,\;\Omega^3_{\;1}=c_1z^{-3}\phi^2,\;\Omega^2_{\;3}=-2c_1z^{-3}\phi^1,\]
which is recognisably type-D, as expected (in this setting, type-D is the condition that the SD Weyl tensor, which is here represented by the real trace-free symmetric matrix $E_{ij}:=\epsilon_i^{\;\;mn}\Omega_{mn\cdot j}$, should be degenerate, in having a repeated eigen-value).

\medskip

We have recovered the expressions in \cite{pb} as deductions from the assumption of Riemannian, Ricci-flat and either Hermitian or one-sided type-D, and we see why there is always a Killing vector preserving the complex structure.

\medskip

Note that
\begin{itemize}
 \item
The holomorphic one-forms are now
\be\label{11}e^1=W^{-1/2}(dt+\omega)+iW^{1/2}dz,\;\;e^2=W^{1/2}e^{u/2}(dx+idy).\ee

\item We can always set $c_1=1$ by a constant rescaling of the metric and redefinition of some coordinates, and this is the value arrived at in \cite{pb2}: we've recovered their expressions exactly.
\item The 2-form $z^{-2}\phi^1$ is closed: this is the rescaling that makes the metric K\"ahler (for the fact that there must be one, see e.g. \cite{dt}).
\end{itemize}
%
 \section{A second Killing vector}
 Suppose there is a second Killing vector, then we may write it as
 \[L=A\partial_t+B\partial_x+C\partial_y+D\partial_z,\]
(with $C$ not as in the previous section).  We start by showing it must have a restricted form, which then allows the Toda field equation to be linearised.

 Write $\L$ for the Lie derivative along $L$, and assume $L$ commutes with $K=\partial_t$ so that
 \[\L(K)=0,\]
 which forces $A,B,C,D$ to be independent of $t$. Also
 \[0=\L(g(K,K))=\L(W^{-1})\mbox{  so that }\L(W)=0.\]
 The curvature is constant along $L$ so that $\L\psi_{ABCD}$ is zero, therefore $\L\psi=0$ (with $\psi$ the scalar in (\ref{1})) and then by the previous section $\L z=0$ and so $D=0$. Also the Lie derivative of $\phi_{AB}$ must be zero (since $-4\psi_{ABCD}=\psi\phi_{(AB}\phi_{CD)}$), and so the complex structure is preserved. From the vanishing of $\L(W)$ we now deduce
 \be\label{b1} \L(u_z)=0.\ee

Since $K, W$ and $J$ are preserved by $L$, we deduce
 \[\mathcal{L}_L(e^1)=0,\]
when, by raising the index, we'll also have
 \[0=\L(W^{1/2}\partial_t+iW^{-1/2}\partial_z)=iW^{-1/2}\L(\partial_z),\]
 so that $\L(\partial_z)=0$ and $A,B,C$ are independent of $z$.

 What can we say about $\mathcal{L}_L(e^2)$? The complex structure is preserved so
 \[\mathcal{L}_L(e^2)=\alpha e^1+\beta e^2\]
 for some $\alpha,\beta$. Already
 \[\mathcal{L}_L(e^1)=0\mbox{  so also  }\mathcal{L}_L(\overline{e}^1)=0\]
 and
 \[\mathcal{L}_L(g(\overline{e}^1,e^2))=0\mbox{  whence  }g(\overline{e}^1,\mathcal{L}_L(e^2))=0,\]
 so $\alpha=0$, and by considering $\mathcal{L}_L(g(\overline{e}^2,e^2))$ we find $\beta$ pure imaginary: $L$ rotates the o.n. basis in the $(\theta^2,\theta^3)$-plane.

 From (\ref{8}) we have the freedom
 \[t\rightarrow t+f(x,y),\;\omega\rightarrow\omega-df,\]
 and under this
 \[L=A\partial_t+B\partial_x+C\partial_y\rightarrow A\partial_t+B(\partial_x+f_x\partial_t)+C(\partial_y+f_y\partial_t),\]
 and in particular
 \[A\rightarrow A+Bf_x+Cf_y,\]
 which we can exploit to set $A=0$.
From
\[0=\L(K)=\L(W^{-1}(dt+\omega)) \mbox{  we deduce  }\L(\omega)=0.\]
Now all that remains of the Killing equation for $L$ is
\be\label{b5}\L(e^u(dx^2+dy^2))=0,\ee
i.e. $L=B(x,y)\partial_x+C(x,y)\partial_y$ is a Killing vector of the 2-metric $h=e^u(dx^2+dy^2)$. We consider this problem in a subsection.
\subsection{Solving (\ref{b5}) for $L$}
We are considering the 2-metric
\[h:=e^u(dx^2+dy^2),\]
with $u(x,y,z)$ which for our problem we can assume analytic (as solutions of the Toda equation must be). We perform the following calculation at a fixed but arbitrary $z$, which we can take to be $z=0$. First calculate the Ricci scalar of $h$:
\[R=-e^{-u}(u_{xx}+u_{yy}).\]
We are requiring the existence of a Killing vector
\[L=B(x,y)\partial_x+C(x,y)\partial_y,\]
which must preserve $R$ and so must be orthogonal to $dR$, so for some $\Omega(x,y)$ we have
\be\label{c1}B=-\Omega R_y,\;\;C=\Omega R_x.\ee
The Killing equations are
\[0=\mathcal{L}_Lh_{ab}=L^c\partial_c h_{ab}+h_{ac}\partial_bL^c+h_{cb}\partial_aL^c.\]
With $ab=12$ this gives
\be\label{c2}B_y+C_x=0,\ee
while with $ab=11$ or $22$
\be\label{c3}B_x=C_y=-\frac12L(u).\ee
Substitute from (\ref{c1}) into (\ref{c2}) and (\ref{c3}) to obtain the following system for $d\Omega$:
\[\Omega_xR_x-\Omega_yR_y=\Omega(R_{yy}-R_{xx})\]
\[\Omega_xR_y+\Omega_yR_x=-2\Omega R_{xy},\]
which algebraically solves to give
\[\frac{\Omega_x}{\Omega}=D^{-1}(R_x(R_{yy}-R_{xx})-2R_yR_{xy}),\]
\[\frac{\Omega_y}{\Omega}=D^{-1}(-R_y(R_{xx}-R_{yy})-2R_xR_{xy}),\]
with $D =(R_x)^2+(R_y)^2=|dR|^2$ (note this is \emph{not} $h(dR,dR)$). This can be simplified by moving some terms to the left since
\[D_x=2R_xR_{xx}+2R_yR_{xy},\;D_y=2R_xR_{xy}+2R_yR_{yy}\]
so that
\[(\log(\Omega D))_x=D^{-1}R_x(R_{xx}+R_{yy}),\]
\[(\log(\Omega D))_y=D^{-1}R_y(R_{xx}+R_{yy}),\]
or
\be\label{c4}d(\log(\Omega D))=D^{-1}\Delta_0 RdR,\ee
with $\Delta_0 R:=R_{xx}+R_{yy}$ (which again is \emph{not} $\Delta_h$). Integrability for (\ref{c4}) is then
\[d\left(D^{-1}\Delta_0 R\right)\wedge dR=0,\]
which is solved by
\[D^{-1}\Delta_0 R=F(R)\]
for some (analytic) $F$. This says
\[\Delta_0 R+F(R)|\nabla R|^2=0\]
so solve $G''/G'=F$ for $G$ to deduce
\[\nabla\cdot(G'\nabla R)=0,\]
and
\[\Delta(G(R))=0,\]
with $\Delta_0$ or $\Delta_h$.

We have {\bf{a necessary condition: some function of $R$ is harmonic.}} Next we exploit the coordinate freedom: with $\zeta=x+iy$ we can make the change
\[\zeta\rightarrow\hat{\zeta}=f(\zeta),\;\hat{x}=\phi(x,y),\;\hat{y}=\psi(x,y),\]
where $\phi,\psi$ are conjugate harmonic functions, if we accompany this by
\[u\rightarrow\hat{u}=u-\log{f'}-\log(\overline{f}'),\]
and then $R$ is unchanged (this is also of course a symmetry of the Toda field equation (\ref{13})). Now if $G(R)$ is harmonic then there is a new coordinate system in which it is $\hat{x}$. Drop the hats then w.l.o.g. $R=R(x)$ and the candidate Killing vector is $\partial_y$. (If preferred we could choose new coordinates so that $R=R(r)$ i.e. radially symmetric at least in some neighbourhood, possibly not including either the origin or a complete circle.)

%
%
%
We have
\be\label{b6} L=\partial_y \mbox{  and  }u_y=0 \mbox{  at  }z=0.\ee
By (\ref{b1}) we shall also have
\[0=\L(u_z)=u_{zy}\]
at $z=0$. Now by uniqueness of solution for the Toda field equation (\ref{13}) we shall have $u_y=0$ for all $z$, and the second Killing vector is, without loss of generality, $L=\partial_y$ everywhere.

\section{Ward's linearisation of the Toda field equation}
In this section, we follow \cite{w1} to relate the $y$-independent Toda field equation:
\be\label{40}u_{xx}+(e^u)_{zz}=0,\ee
to the axisymmetric Laplace equation in cylindrical polars:
\be\label{41}V_{ZZ}+R^{-1}(RV_R)_R=0.\ee
This will lead to a solution of the Ricci-flat equations for the general one-sided type-D metric with an extra symmetry in terms of a solution of the axisymmetric Laplace equation in three dimensions.

To see how $u$ and $V$ are related, and following \cite{w1}, set
\be\label{42}x=V_Z,\;z=\frac12RV_R,\;u=\log(R^2/4),\ee
(so we need to suppose that $V_R,V_Z$ are not constant). We calculate the Jacobian matrix
\[\frac{\partial(x,z)}{\partial(R,Z)}=
\left(\begin{array}{cc}
 V_{ZR}&-\frac12 RV_{ZZ}\\
            V_{ZZ} & \frac12 RV_{RZ}\\
 \end{array}\right)\]
with the aid of (\ref{41}), and the inverse is
\be\label{jac}\frac{\partial(R,Z)}{\partial(x,z)}=\Delta^{-1}
\left(\begin{array}{cc}
 \frac12 RV_{ZR}&\frac12 RV_{ZZ}\\
            -V_{ZZ} & V_{RZ}\\
 \end{array}\right)\ee
 with $\Delta=\frac12 R((V_{RZ})^2+(V_{ZZ})^2)$ (since $V_R,V_Z$ are not constant, $\Delta$ is nonzero). In particular therefore
\[R_x=\frac12\Delta^{-1}RV_{RZ},\;\;R_z=-\Delta^{-1}V_{ZZ},\]
so with $u=\log(R^2/4)$ as in (\ref{42}) we deduce
\be\label{43} u_x=\frac{2}{R}R_x=\Delta^{-1}V_{RZ}=Z_z,\;e^u u_z=-\frac{R}{2}\Delta^{-1}V_{ZZ}=-Z_x,\ee
and by cross-differentiating, we see that $u$ satisfies (\ref{40}).

With $u_y=0$, $d\omega$ as in (\ref{14}) subject to (\ref{15}) becomes
\[d\omega=-W_xdy\wedge dz-e^u(W_z-2c_1\frac{W^2}{z^2})dx\wedge dy=d(Fdy),\]
with
\[F_z=W_x=-\frac{z^2u_{xz}}{2c_1},\;\;F_x=-e^u(W_z-2c_1\frac{W^2}{z^2})=\frac{1}{2c_1}e^u(-z^2(u_{zz}+(u_z)^2)+2zu_z-2). \]
 To integrate for $F$ we need the function $H$ conjugate to $V$. From (\ref{41}) this satisfies
 \[H_R=RV_Z,\;\;H_Z=-RV_R,\]
 and then one verifies that
 \[F=\frac{1}{2c_1}(z^2u_x-\frac12 xR^2+H).\]
Now transform the metric. First note that
\[dz^2+e^udx^2=\left(\frac12(RV_R)_RdR+\frac12 RV_{RZ}dZ\right)^2+\frac{R^2}{4}\left(V_{RZ}dR+V_{ZZ}dZ\right)^2\]
\[=\frac{R^2}{4}\left((-V_{ZZ}dR+V_{RZ}dZ)^2+(V_{RZ}dR+V_{ZZ}dZ)^2\right)\]
\[=\frac12R\Delta(dR^2+dZ^2),\]
with the aid of (\ref{41}) and the definition of $\Delta$. Now the metric is
\[g=W^{-1}(dt+\omega)^2+W(dz^2+e^u(dx^2+dy^2))\]
\[=W^{-1}(dt+Fdy)^2+We^udy^2+\frac12WR\Delta(dR^2+dZ^2),\]
=\[\left(\begin{array}{cc}
 dt&dy\\
 \end{array}\right)\left(\begin{array}{cc}
 W^{-1}&FW^{-1}\\
            FW^{-1} & F^2W^{-1}+We^u\\
 \end{array}\right)\left(\begin{array}{c}
 dt\\
            dy\\
 \end{array}\right) +\Omega^2(dR^2+dZ^2) \]
which is the canonical form for a Ricci-flat metric with two commuting Killing vectors. Note that the determinant of the matrix of Killing vector contractions is $e^u=R^2/4$ so that $R$ (up to a constant factor) is the standard radial coordinate, and then $Z$ is its harmonic conjugate. The metric components can be given explicitly in terms of $V,R,Z$ by noting
\[W=\frac{V_R}{2c_1}\left(\frac{R((V_{RZ})^2+(V_{ZZ})^2)+V_RV_{ZZ}}{((V_{RZ})^2+(V_{ZZ})^2)}\right),\]
\[F=\frac{1}{2c_1}\left(H+\frac{R(V_R)^2V_{RZ}-R^2V_Z((V_{RZ})^2+(V_{ZZ})^2)}{2((V_{RZ})^2+(V_{ZZ})^2)}\right),\]
\[\Omega^2=\frac{1}{8c_1}R^2V_R\left(R((V_{RZ})^2+(V_{ZZ})^2)+V_RV_{ZZ}\right).\]
 Note that:
 \begin{itemize}
  \item We have linearised the field equations for a Ricci-flat metric with two commuting symmetries. These field equations are already known to be linear if one of the symmetries is hypersurface orthogonal. We can be sure that we haven't inadvertently reduced to this case by looking at the examples which follow, specifically the Riemannian Kerr solution which does not in general admit a hypersurface orthogonal Killing vector.
 \item
We have an expression for the metric given harmonic $V(R,Z)$. We could think about getting back to $V$ starting from $g$. One route is to set $Q=RV_R$ when
 \[8c_1\Omega^2=Q((Q_Z)^2+(Q_R)^2-R^{-1}QQ_R)\]
 which needs to be solved for $Q$ given $\Omega$, with $Q$ also subject to
 \[Q_{RR}-R^{-1}Q_R+Q_{ZZ}=0.\]
 \end{itemize}
%
\subsection{A Corollary: linearising the SD Einstein equations with two commuting symmetries}
As a corollary to the previous section, we recall that the general solution to the (four-dimensional) SD (or ASD) Einstein equations with a symmetry was given in \cite{p} (see also \cite{t1}, \cite{t2}) and it also depends on a solution to the Toda field equation. Thus with a second symmetry, the Ward transformation can be applied again to give the general solution in this case in terms of an axisymmetric solution of the Laplace equation.

Recall, from \cite{t1}, the metric in this case is
\be\label{444}
g=\frac{1}{Pz^2}(dt+\omega)^2+\frac{P}{z^2}(dz^2+e^u(dx^2+dy^2)),\ee
with
\[u_{xx}+u_{yy}+(e^u)_{zz}=0,\]
\[2\Lambda P=zu_z-2,\]
where $\Lambda$ is proportional to the (constant) Ricci scalar, and
\[d\omega=-P_xdy\wedge dz-P_ydz\wedge dx-(Pe^u)_zdx\wedge dy.\]
If we add a second symmetry then the argument goes through as before, and without loss of generality we can suppose $u_y=0$ and the second symmetry is $L=\partial_y$. We can solve for $\omega$:
\[\omega=Fdy,\;\;F=\frac{1}{2\Lambda}(zu_x-Q),\]
with $Q$ conjugate to $u$ in the sense
\[u_x=Q_z,\;\;(e^u)_z=-Q_x.\]
We follow the Ward transformation as before, set:
\[x=V_Z,\;z=\frac12RV_R,\;\;u=\log(R^2/4),\]
then as before
\[dz^2+e^udx^2=\frac12R\Delta(dR^2+dZ^2),\]
with
\[\Delta=\frac12R((V_{RZ})^2+(V_{ZZ})^2).\]
The metric becomes
\be\label{45}
g=\frac{8\Lambda\Delta}{R^2V_R^2(2\Delta+V_RV_{ZZ})}(dt+Fdy)^2+\frac{(2\Delta+V_RV_{ZZ})}{2\Lambda\Delta V_R^2}dy^2+\frac{(2\Delta+V_RV_{ZZ})}{RV_R^2}(dR^2+dZ^2).\ee
For $F$ we note from (\ref{43}) that $Q=Z$ and from the Jacobian matrix (\ref{jac})
\[zu_x=\frac{1}{2\Delta}RV_RV_{ZZ},\]
so that
\[F=\frac{1}{2\Lambda}(\frac{1}{2\Delta}RV_RV_{ZZ}-Z).\]
This then is the general SD Einstein metric with $\Lambda\neq 0$ and two commuting symmetries, written in terms of a solution $V$ of the axisymmetric Laplace equation (\ref{41}).
We'll give an example in Section 7.
\section{One-sided type-D Einstein metrics}
It's natural to ask whether the addition of a non-zero scalar curvature interferes with the integrability of the field equations: for ASD vacuum with a symmetry it does not while for vacuum with two commuting Killing vectors it does. In the case considered in this section it does -- we can reduce the field equations to a modified Toda field equation  but not an integrable one.\footnote{This class of metrics was considered in \cite{p2} but it is not straightforward to compare results obtained.}

To see this, suppose we still have (\ref{1}) but with a nonzero $\Lambda$, in NP conventions. Then (\ref{2}) still defines a Killing spinor (the argument in \cite{pw} for its existence uses the contracted Bianchi identity, which is unchanged in this case). We need to assume that $\psi$ is not constant as we shall want to use it as a coordinate. By making this assumption we are eliminating some cases: if $\psi$ is constant then $K^a$ is zero and $\omega_{AB}$ defines a K\"ahler form, thus we are excluding Einstein-K\"ahler metrics. Equation (\ref{3}) still defines a Killing vector $K^a$, but (\ref{4}) is changed to
\be\label{46}
\nabla_{AA'}K_{BB'}=-\frac{1}{6}(\psi+12\Lambda)\omega_{AB}\epsilon_{A'B'}+\chi_{A'B'}\epsilon_{AB},\ee
with a corresponding change in (\ref{10}). With $d\omega$ as in (\ref{14}) we find $\alpha$ and $\beta$ as in (\ref{15}) unchanged but
\be\label{46}
\gamma=-e^u(W_z+\frac{2}{3}(\psi+12\Lambda)\psi^{-1/3}).\ee
We may introduce $z=-\psi^{-1/3}$ as before and then it is convenient to introduce
\be\label{47}
g(z)=\frac{1}{3}\psi^{-1/3}(\psi+12\Lambda)=\frac{12\Lambda z^3-1}{3z^2}.\ee
The connection coefficients $\alpha^i_{\;j}$ take the same form with $G$ and $H$ unchanged but with
\be\label{48}
C=W^{1/2}g-\frac12W^{-1/2}u_z,\ee
and
\be\label{49}
E=gW^{1/2}.\ee
We calculate the curvature forms $\Omega^i_{\;j}$ but this time with $\epsilon_{ijk}\Omega_{ij.k}=24\Lambda$. From $\Omega^1_{\;2}+i\Omega^1_{\;3}$ we find
\[gW=-\frac12u_z+h(z)\]
with $h'=h^2$ so that $h=-(z-c)^{-1}$ for some $c$. Then $\Omega^2_{\;3}$ leads to $c=0$ and the modified Toda equation
\be\label{49a}
u_{xx}+u_{yy}+(e^u)_{zz}+e^u(A(z)u_z+B(z))=0,\ee
for $u$, with
\[A=\frac{72\Lambda z^2}{1-12\Lambda z^3},\;\;B=-\frac{144\Lambda z}{1-12\Lambda z^3}.\]
One can now verify that the expression for $d\omega$ is consistent (i.e. that $d(d\omega)=0$). One can also check that (\ref{49a}) cannot just be transformed back into the Toda equation by the simple change of variable:
\be\label{mt2}z\rightarrow Z=f(z),\;\;u\rightarrow U=u+F(z).\ee
This in fact shows that (\ref{49a}) is not integrable: it is known (\cite{F}) that the only modifications of Toda of the form of (\ref{49a}) which are integrable are those obtained by the transformation (\ref{mt2}). We can check the expressions for $A(z)$ and $B(z)$ by obtaining the known one-sided type D Einstein space which is the Schwarzschild-de Sitter metric. We do this in the final section, with other examples. In the next section we dispose of an apparent puzzle concerning the one-sided type-D vacuum and Einstein metrics.
\section{An apparent puzzle}
From the assumption of one-sided type-D-ness and either vacuum or Einstein, we have been able to reduce the Einstein equations to one of two PDEs for a single function $u$. It is striking that at no stage do we consider the ASD Weyl spinor, only ever the SD Weyl spinor, but the ASD Weyl spinor is evidently uniquely defined from the metric. This could be contrasted with linear (gravity) theory in which there are degrees of freedom and therefore free data to be specified for both SD and ASD linearised Weyl spinors. One way to explain this apparent puzzle is to point to the Killing vector arising from the Killing spinor, which necessarily satisfies the identity
\be\label{k1}
\nabla_a\nabla_bK_c=R_{bcad}K^d,
\ee
a relation that one doesn't have in linear theory but which, in the full theory, clearly connects the ASD and SD Weyl spinors. Dealing first with one-sided type-D vacuum solutions, we have the Killing spinor $\omega_{AB}$ from (\ref{2}), defining the Killing vector $K^a$ via (\ref{3}), which in turn satisfies (\ref{4}). If we contract (\ref{4}) with $K^b$ we obtain
\[\frac12\nabla_a W^{-1}= K^b\nabla_aK_b =-\frac{1}{6}\psi\omega_{AB}K^B_{\;A'}-\chi_{A'B'}K^{B'}_{\;A}\]
which we may rearrange as
\[\chi_{A'B'}K^{B'}_{\;A}=\frac{1}{36}\psi^{-2/3}\nabla_a\psi-\frac12\nabla_aW^{-1}=\nabla_aQ\]
where we introduce $Q=-\frac12W^{-1}-\frac{1}{12}z^{-1}$. Multiplying by $K^A_{\;A'}$ gives an expression for $\chi_{A'B'}$ as
\[\chi_{A'B'}=WK^B_{\;A'}\nabla_{BB'}Q\]
which is therefore known explicitly when we know $u$. This in turn is a potential for the SD Weyl spinor via (\ref{k4}) so that, as must be the case, knowledge of $u$ fixes both the ASD and the SD Weyl spinor -- there is no more free data.

\section{Examples}

\begin{itemize}
  \item  Starting with one-sided type-D vacua, flat space is \emph{not} an example of the construction, as we're assuming the SD Weyl spinor isn't zero. However, we could have chosen the zero solution of the Toda equation, when, taking $c_1=1$ for simplicity,
\[u=0,\;\;W=z,\;\;\omega=xdy,\]
and the metric is
\[g=\frac{1}{z}(dt+xdy)^2+z(dz^2+dx^2+dy^2),\]
which is recognisably the Gibbons-Hawking metric with potential $z$ (see e.g. \cite{GH}). In particular, this metric is hyper-K\"ahler with the other orientation so that the primed Weyl spinor is zero. With $T=\frac{2}{3}z^{3/2}$, it can be written
\[g=dT^2+\left(\frac{3T}{2}\right)^{3/2}(dx^2+dy^2)+\left(\frac{2}{3T}\right)^{3/2}(dt+xdy)^2,\]
 which makes the isometry group manifest: this is LRS Bianchi-type II. Because $u=0$, this doesn't have a Ward form.
 \item There are separable solutions of the Toda equation (\ref{13}) in the sense $u=f(x,y)+g(z)$ (see e.g. \cite{t3}) some of which can be written
  \[u=-2\log(1+x^2+y^2)+\log(4(z^2+2mz+a)),\]
  for real constants $a,m$, when
  \[W=\frac{z(a+mz)}{z^2+2mz+a},\;\;\omega=-a\cos\theta d\phi,\]
  and we've introduced polar coordinates by $\zeta=\tan(\theta/2)e^{i\phi}$. The metric can be written
  \be\label{51}g=\frac{(z^2+2mz+a)}{z(a+mz)}(dt-a\cos\theta d\phi)^2+\frac{z(a+mz)}{z^2+2mz+a}dz^2+z(a+mz)(d\theta^2+\sin^2\theta d\phi^2),\ee
  which has LRS Bianchi-type-IX form. When $a=0$ it is the Riemannian Schwarzschild solution; with $a=m^2, n=-m^{3/2}/2$ it is the self-dual Taub-NUT metric as given in equation (3.9) of \cite{GH}; with $m=0$ it is the Eguchi-Hanson metric as in (3.20) of \cite{GH} but with $a^4$ there replaced by $-16a^3$ from here (in particular it \emph{isn't} the Riemannian Kerr solution).

  There will be more, probably unfamiliar, one-sided type-D metrics determined by the `quadric ansatz' \cite{t3} for solutions of the Toda equation.
  \item To illustrate the Ward transformation on a one-sided type-D vacuum with an extra symmetry, we consider the particular separable solution of the Toda equation given by
  \[u=2\log\mbox{sech}\, x+\log(z^2+2mz+a),\]
  when
  \[W=\frac{z(mz+a)}{z^2+2mz+a},\;\omega=a\mbox{tanh}xdy,\]
  and with $\cos\theta=\tanh x,\phi=-y$ we arrive at the metric in (\ref{51}) again, but this time with $u$ such that $u_y=0$. For simplicity put $a=m^2$, and then
  \[R=2(z+m)\mbox{sech}\,x,\;Z=-2(z+m)\tanh x,\]
  and
  \[V=-2m\log R+(R^2+Z^2)^{1/2}-Z\tanh^{-1}\left(\frac{Z}{(R^2+Z^2)^{1/2}}\right),\]
  which one verifies is harmonic. (In spherical polars the terms independent of $m$ are
  \[V=r(1-\cos\theta\log\cot(\theta/2).)\]
  \item For the transformation (\ref{42}) to be nontrivial, we need $V_R$ and $V_Z$ to be nonconstant, so for an example with a simple $V$ consider
  \[V=R^2-2Z^2.\]
  Then
  \[x=V_Z=-4Z,\;\;z=\frac12RV_R=R^2,\;\;u=\log(R^2/4),\]
  and we have the simple solution $u=\log(z/4)$ of (\ref{40}). After a change of variable, the metric (\ref{8}) becomes the LRS Riemannian Kasner solution:
  \[g=dT^2+T^{-2/3}dU^2+T^{4/3}(dX^2+dY^2).\]
  \item {\bf{Riemannian Kerr:}} We start with the NP tetrad (of vectors) tied to the Principal Null Directions for Lorentzian Kerr as given in \cite{an}, (24)--(26), lower to one-forms and transform to Boyer-Lindquist coordinates via (35) of \cite{an} to obtain
  \[\ell=dt-\frac{\Sigma^2}{\tD}dr-a\sin^2\theta d\phi,\;n=\frac{\tD}{2\Sigma^2}(dt-a\sin^2\theta d\phi)+\frac12dr,\]
  \[m=\frac{1}{\sqrt{2}\Gamma}(ia\sin\theta dt-\Sigma^2d\theta-i(r^2+a^2)\sin\theta d\phi),\]
  where
  \[\Sigma^2=r^2+a^2\cos^2\theta,\;\;\tD=r^2-2mr+a^2,\;\;\Gamma=r+ia\cos\theta.\]
  In this form we can analytically continue to Riemannian signature by changing $(t,a)$ to $(it,ia)$. It's convenient to boost and rotate the basis a little to arrive at
  \[L=(-i\tth)\ell=\tth(dt-a\sin^2\theta d\phi)+\frac{i}{2\tth}dr,\;\;N=\overline{L},\]
  and
  \[M=\frac{1}{\sqrt{2}\Sigma}(-\Sigma^2d\theta+i\sin\theta(adt+(r^2-a^2)d\phi))\]
  where now
  \[\tth=\frac{\sqrt{\tD}}{\Sigma\sqrt{2}},\;\;\tD=r^2-2mr-a^2,\;\;\Sigma^2=r^2-a^2\cos^2\theta.\]
  Since both Weyl spinors are type-D we have a choice of complex structures, both of them integrable: one, say $J_1$, has $L,M$ as holomorphic one-forms and the other, $J_2$, has $L,\overline{M}$. We lower the Killing vector $K=\partial_t$ and take its exterior derivative to obtain
  \[dK=iX(L\wedge\overline{L}+M\wedge\overline{M})+iY(L\wedge\overline{L}-M\wedge\overline{M})\]
  with
  \[X=-m(r-a\cos\theta)^{-2},\;\;Y=-m(r+a\cos\theta)^{-2},\]
  so if we stick with $J_1$ then $i(L\wedge\overline{L}+M\wedge\overline{M})$ is the 2-form corresponding to the Killing spinor under consideration, the scalar $\psi$ is a multiple of $(r-a\cos\theta)^{-3}$ and the coordinate $z$ is a multiple of $r-a\cos\theta$. Comparing the Riemannian Kerr metric, which is now
  \[g=2\tth^2(dt-a\sin^2\theta d\phi)^2 +\frac{dr^2}{2\tth^2}+\Sigma^2d\theta^2+\frac{\sin^2\theta}{\Sigma^2}(adt+(r^2-a^2)d\phi)^2\]
  with the metric form (\ref{8}) and using $z=r-a\cos\theta$, we are led to
  \[e^ud\zeta d\overline{\zeta}=\tD\sin^2\theta\left((a\tD^{-1}dr-\csc\theta d\theta)^2+d\phi^2\right),\]
  when a choice for $\zeta$ is
  \[\zeta=x+iy=\log\left(\left(\frac{r-m-b}{r-m+b}\right)^{-a/2b}\tan(\theta/2)e^{i\phi}\right),\]
  where $b^2=a^2+m^2$, so in particular $y=\phi$, and then $u$ is given by
  \[e^u=(r^2-2mr-a^2)\sin^2\theta,\]
  in agreement with (\ref{42}) (as this is the determinant of the $(\phi,t)$-part of the metric). It is straightforward to verify that
  \[u_x=Z_z,\;\;e^uu_z=-Z_x,\]
  with $Z=2(r-m)\cos\theta$, so that $u$ does satisfy (\ref{40}) but we can't obtain $u(x,z)$ explicitly. From (\ref{42}) we have
  \[R^2=4e^u=4(r^2-2mr-a^2)\sin^2\theta,\]
  together with
  \[Z=2(r-m)\cos\theta,\]

  so that $(r,\theta)$ are ellipsoidal coordinates in the $(R,Z)$ plane. It is straightforward to obtain $V(r,\theta)$: we find
  \[V=2(r-a\cos\theta)+2((r-m)\cos\theta-a)\log\tan(\theta/2)+2m\log\sin\theta\]\[+((m+b)-\frac{a}{b}(r-m)\cos\theta)\log(r-m-b)+((m-b)+\frac{a}{b}(r-m)\cos\theta)\log(r-m+b).\]
  This isn't simple for $V$ in terms of $R,Z$ (or $x,z$).
  \item We can give a simple example of the construction described in the Corollary (Section 4.1) by again taking $V=R^2-2Z^2$. With $\Lambda=2$ for convenience (which leads to Ricci scalar equal to $-24$), and $Y=y/4$, the metric turns out to be
  \be\label{52}
  g=\frac{4}{R^4}(dt-ZdY)^2+\frac{1}{ R^2}(dR^2+dY^2+dZ^2).\ee
  This is easily to be seen to be ASD Einstein-K\"ahler with constant holomorphic sectional curvature and negative Ricci scalar, so it must be the Bergman metric.

\item As an example of one-sided type-D Einstein from Section 5, we note that (\ref{49a}) has solutions of the form
\[u=-2\log(1+k(x^2+y^2))+F(z),\]
with $e^F$ a 2-parameter family of quartic polynomials in $z$:
\[e^F=c_2(1+24\Lambda z^3)+c_3z(1+6\Lambda z^3)+4kz^2,\]
 and these include the Schwarzschild-de Sitter metric if $k=1, c_2=0$, and $c_3$ is negative and related to the mass parameter. This calculation also confirms the expressions for $A$ and $B$ in (\ref{49a}).

 \end{itemize}

\end{document}